\documentstyle[12pt]{article}
\begin{document} 
\thispagestyle{empty} 
\begin{flushright}
UA/NPPS-6-04\\
\end{flushright}

\vspace{2cm}
\begin{center}
{\large{\bf
LOCATING THE QCD CRITICAL POINT \\
IN THE PHASE DIAGRAM\\}}
\vspace{1cm} 
{\large N. G. Antoniou, F. K. Diakonos and A. S. Kapoyannis}\\ 
\smallskip 
{\it Department of Physics, University of Athens, 15771 Athens, Greece}\\ 

\vspace{1cm}

Presented at the X International Workshop on Multiparticle Production\\
``Correlations and Fluctuations on QCD'', \\
8-15 June 2002, Istron Bay, Ag. Nikolaos, Crete, GREECE

\end{center}
\vspace{0.5cm}
\begin{abstract}
It is shown that the hadronic matter formed at high temperatures, according
to the prescription of the statistical bootstrap principle, develops a
critical point at nonzero baryon chemical potential, associated with the end
point of a first-order, quark-hadron phase-transition line. The location of
the critical point is evaluated as a function of the MIT bag constant.
\end{abstract}

\vspace{3cm}


\newpage
\setcounter{page}{1}

{\bf 1 Introduction}

Quantum Chromodynamics is unquestionably the microscopic theory of strong
interactions and offers an accurate description of quark-gluon matter. The
formation of hadronic matter is still an open problem in the context of QCD.
This theory predicts however the existence of a critical point at non zero
baryon chemical potential, which is the end point of a quark-hadron critical
line of first order [1]. This singularity is associated with the formation of
hadronic matter at high temperatures and its location in the QCD phase
diagram is of primary importance.

On the other hand the hadronic side of matter can be treated as a thermally
and chemically equilibrated gas. The inclusion of {\em interactions} among
hadrons is crucial in order to reveal the possibility of a phase transition. 
A model that allows for the thermodynamical description of interacting
hadrons is the Statistical Bootstrap Model (SBM), which was first developed
by Hagedorn [2-5]. In what follows we investigate the possibility of the
formation of a critical point within the framework of the statistical
bootstrap hypothesis.

\vspace{2cm}
{\bf 2 The hadronic matter}

The SBM is based on the hypothesis that the strong interactions can be
simulated by the presence of hadronic clusters.
In the context of SBM the strongly interacting hadron gas is replaced by a
non-interacting infinite-component cluster gas. 
The hadronic states of clusters are listed in a mass spectrum
$\tilde{\rho}$, so that $\tilde{\rho}dm$ represents the number of hadronic
states in the mass interval $\{m,m+dm\}$. The mass spectrum can be
evaluated if the clusters, as well as, their constituents are treated on the
same footing by introducing an integral {\it bootstrap} equation (BE).
In the bootstrap logic clusters are composed
of clusters described by the same mass spectrum. This scheme proceeds until
clusters are reached that their constituents cannot be divided further. These
constituents are the {\it input} hadrons and the known hadronic particles 
belong to this category. The BE leads to the adoption of an asymptotic mass
spectrum of the form [7]
\begin{equation} 
\tilde{\rho}(m^2,\{\lambda\})\stackrel{m\rightarrow\infty} 
{\longrightarrow} 
2C(\{\lambda\})m^{-\alpha} \exp [m\beta^*(\{\lambda\})]\;. 
\end{equation} 
The underlying feature of SBM is that the mass spectrum rises
{\it exponentially}, as m tends to infinity.
$\beta^*$ is the inverse maximum temperature allowed for hadronic matter 
and depends on the existing fugacities $\{\lambda\}$. $\alpha$ is an exponent
which can be adjusted to different values allowing for different versions of
the model.

The manipulation of the bootstrap equation can be significantly simplified 
through suitable Laplace transformations. The Laplace transformed mass
spectrum leads to the introduction of the quantity $G(\beta,\{\lambda\})$.
The same transformation can be carried out to the input term of SBM, leading
to the quantity $\varphi(\beta,\{\lambda\})$. Then the BE can be expressed as
\begin{equation} 
\varphi(\beta,\{\lambda\})=2G(\beta,\{\lambda\}) 
-\exp [G(\beta,\{\lambda\})]+1\;. 
\end{equation}
The above BE exhibits a singularity at  
\begin{equation} 
\varphi(\beta,\{\lambda\})=\ln4-1\;. 
\end{equation}
The last equation imposes a constraint among the thermodynamic variables
which represent the {\em boundaries} of the hadronic phase. Hadronic matter
can exist in all states represented by variables that lead to a real solution
of the BE or equivalently in all states for which temperatures and fugacities
lead to
\begin{equation} 
\varphi(\beta,\{\lambda\})\le\ln4-1\;. 
\end{equation}

In the general form of SBM the following four improvements can be 
made which allow for a better description of hadronic matter:

1) The inclusion of all the known hadrons with masses up to 2400 MeV in the
input term of the BE and also inclusion of strange hadrons. This leads to the
introduction of the strangeness fugacity $\lambda_s$ in the set of fugacities
[6,7]. Another fugacity which is useful for the analysis of the experimental
data in heavy ion collisions is $\gamma_s$. This fugacity allows for partial
strangeness equilibrium and can also be included in the set of fugacities of 
SBM [8].

2) Different fugacities can be introduced for $u$ and $d$ quarks. In this way
the thermodynamic description of systems which are not isospin symmetric
becomes possible. Such systems can emerge from the collision of nuclei with
different number of protons and neutrons [9].

3) The choice of the exponent $\alpha$ in (1) has important consequences,
since every choice leads to a different physical behaviour of the system. The
usual SBM choice was $\alpha=2$, but more advantageous is the choice
$\alpha=4$. With this choice a better physical behaviour is achieved as the 
system approaches the hadronic boundaries. Quantities like 
pressure, baryon density and energy density, even for 
point-like particles, no longer tend to infinity as the system 
tends to the bootstrap singularity.
It also allows for the bootstrap singularity to be reached in the 
thermodynamic limit [10], a necessity imposed by the Lee-Yang theory.
Another point in favour of the choice $\alpha=4$ comes from the extension of
SBM to include strangeness [6,7]. 
The strange chemical potential equals zero in the quark-gluon phase. With
this particular choice of $\alpha$, $\mu_s$ acquires smaller positive values
as the hadronic boundaries are approached.
After choosing $\alpha=4$ the partition function can be written down and
for point-like particles it assumes the form
\begin{equation} 
\ln Z_{p\;SBM}(V,\beta,\{\lambda\})= 
\frac{4BV}{\beta^3}\int_{\beta}^{\infty} x^3 G(x,\{\lambda\})dx \equiv
Vf_{SBM}(\beta,\{\lambda\})\;,
\end{equation} 
where $B$ is the energy density of the vacuum (bag constant) and it is the
only free parameter of SBM which is left after fixing $\alpha=4$ [6,7].

4) The contributions due to the finite size of hadrons, accounting for the
repulsive interaction among hadrons, can be introduced via a Van der Waals
treatment of the volume. The negative contributions to the volume can be
avoided if the following grand canonical pressure partition function is used
\begin{equation} 
\pi(\xi,\beta,\{\lambda\})=\frac{1}{\xi-f(\beta+\xi/4B,\{\lambda\})}\;,
\end{equation} 
where $\xi$ is the Laplace conjugate variable of the volume.
All values of $\xi$ are allowed if Gaussian regularization 
is performed [11]. The value $\xi=0$ corresponds to a system without external
forces [10,11] and it will be used throughout our calculations.
With the use of (6) and the SBM point particle partition function (5)
one obtains
\begin{equation} 
\nu_{HG}(\xi,\beta,\{\lambda\})=\lambda
\frac{\frac{\partial f(\beta+\xi/4B,\{\lambda\})}{\partial \lambda}}
{{\scriptstyle 1-}\frac{1}{4B}
\frac{\partial f(\beta+\xi/4B,\{\lambda\})}{\partial \beta}}\;,
\end{equation} 
where $\lambda$ is the fugacity corresponding to the particular density, and
\begin{equation}
P_{HG}(\xi,\beta,\{\lambda\})=\frac{1}{\beta}
\frac{{\scriptstyle f(\beta+\xi/4B,\{\lambda\})-}\frac{\xi}{4B}
\frac{\partial f(\beta+\xi/4B,\{\lambda\})}{\partial \beta}}
{{\scriptstyle 1-}\frac{1}{4B}\frac{\partial f(\beta+\xi/4B,\{\lambda\})}{\partial \beta}}\;.
\end{equation} 

The dependence of the pressure on the volume can be recovered if
for a given set of parameters $\xi$, $\beta$, $\{\lambda\}$ the density
$\nu_b$ of the conserved baryon number $<b>$ is calculated. Then the volume
would be retrieved through the relation
\begin{equation}
<V>=\frac{<b>}{\nu_b}\;.
\end{equation} 

By using the SBM with all the above improvements the possibility of a phase
transition of hadronic matter can be traced. The study of the pressure-volume
isotherm curve is then necessary. 
When this curve is calculated one important feature of SBM is revealed. This
curve has a part (near the boundaries of the hadronic domain) where pressure 
decreases while volume decreases also (see Fig.1). This behaviour is due to
the formation of bigger and bigger clusters as the system tends to its
boundaries. Such a behaviour is a signal of a {\em first order phase
transition} which in turn is connected with the need of a {\em Maxwell
construction}.

If on the contrary the interaction included in SBM is not used then no such
behaviour is exhibited. This can be verified if the Ideal Hadron Gas model is
used. Then for this model the equation that corresponds to Eq. (5) is
\begin{equation} 
f_{p\;IHG}(\beta,\{\lambda\}) \equiv
\frac{\ln Z_{p\;IHG}(V,\beta,\{\lambda\})}{V}=\frac{1}{2\pi^2\beta}
\sum_{\rm a} [\lambda_{\rm a}(\{\lambda\})+\lambda_{\rm a}(\{\lambda\})^{-1}]
\sum_i g_{{\rm a}i} m_{{\rm a}i} K_2 (\beta m_{{\rm a}i})\;, 
\end{equation}
where $g_{{\rm a}i}$ are degeneracy factors due to spin and isospin and
{\rm a} runs to all hadronic families.
This function can be used in eq. (6) to calculate the Ideal Hadron Gas (IHG)
pressure partition function in order to include Van der Waals volume
corrections. The result is that the pressure is always found to increase as
volume decreases, for constant temperature, allowing for no possibility of a
phase transition.

The comparison of SBM with the IHG (with volume corrections) is displayed in
Figure 1, where $\nu_0$ is the normal nuclear density $\nu_0=0.14\;fm^{-3}$.
In both cases (SBM or IHG) the constraints $<S>=0$ (zero strangeness)
and $<b>=2<Q>$ (isospin symmetric system, i.e. the net number of $u$ and $d$
quarks are equal) have been imposed. Also strangeness is fully equilibrated
which accounts to setting $\gamma_s=1$.

\vspace{2cm}
{\bf 3 The quark-gluon matter}

We may now proceed to the 
thermodynamical description of the quark-gluon phase. 
The grand canonical partition function of a system 
containing only $u$ and $d$ massless quarks and gluons is [13]
\begin{eqnarray}
\ln Z_{QGP}(V,\beta,\lambda_q)=&&
\hspace{-0.5cm}\underbrace{\frac{gV}{6\pi^2}\beta^{-3}
\left[\left(1-\frac{2a_s}{\pi}\right)
\left(\frac{1}{4}\ln^4 \lambda_q+\frac{\pi^2}{2}\ln^2 \lambda_q \right)+
\left(1-\frac{50a_s}{21\pi}\right)\frac{7\pi^4}{60}\right]}_
{\rm quark\;term} \nonumber \\
&&\hspace{-0.5cm}+\underbrace{V\frac{8\pi^2}{45}\beta^{-3}\left(1-\frac{15a_s}{4\pi}\right)}
_{\rm gluon\;term}-\underbrace{\beta BV}_{\rm vacuum\;term}\;.
\end{eqnarray}
This partition function is calculated to first order in the QCD running
coupling constant $a_s$. The fugacity $\lambda_q$ is related to both $u$ and
$d$ quarks. $B$ is again the MIT bag constant and $g$ equals to the product
of spin states, colours and flavours available in the system,
$g=N_sN_cN_f=12$. Using this partition function the QGP baryon density and
pressure can be calculated through the relations
\begin{equation}
\nu_{b\;QGP}(\beta,\lambda_q)=
\frac{2}{\pi^2}\beta^{-3}
\left(1-\frac{2a_s}{\pi}\right)
\left(\frac{1}{3}\ln^3 \lambda_q+\frac{\pi^2}{3}\ln \lambda_q \right)\;
\end{equation}

\begin{eqnarray}
P_{QGP}(\beta,\lambda_q)=&&\hspace{-0.5cm}
\frac{2}{\pi^2}\beta^{-4}
\left[\left(1-\frac{2a_s}{\pi}\right)
\left(\frac{1}{4}\ln^4 \lambda_q+\frac{\pi^2}{2}\ln^2 \lambda_q \right)+
\left(1-\frac{50a_s}{21\pi}\right)\frac{7\pi^4}{60}\right] \nonumber \\ 
&&\hspace{-0.5cm}+\frac{8\pi^2}{45}\beta^{-4}\left(1-\frac{15a_s}{4\pi}\right)-B\;.
\end{eqnarray}

If the strange quarks are also included, the quarks assume their
current masses and $a_s=0$, then the following partition function can be
used.
\begin{eqnarray}
\ln Z_{QGP}(V,\beta,\lambda_u,\lambda_d)=&&\hspace{-0.5cm}
\underbrace{\frac{N_s N_c V}{6\pi^2}
\beta\sum_i \int_0^{\infty} \frac{p^4}{\sqrt{p^2+m_i^2}}
\frac{1}{e^{\beta \sqrt{p^2+m_i^2}}\lambda_i^{-1}+1}dp}_{\rm quark\;term} \nonumber \\
&&\hspace{-0.5cm}+\underbrace{V\frac{8\pi^2}{45}\beta^{-3}}_{\rm gluon\;term}-
\underbrace{\beta BV}_{\rm vacuum\;term}\;.
\end{eqnarray}
The index $i$ runs to all quarks and antiquarks. The current
masses are taken $m_u=5.6$ MeV, $m_d=9.9$ MeV and $m_s=199$ MeV [14].
The fugacities are $\lambda_{\bar{u}}=\lambda_u^{-1}$,
$\lambda_{\bar{d}}=\lambda_d^{-1}$ and $\lambda_{\bar{s}}=\lambda_s^{-1}=1$
(since strangeness is set to zero). The baryon density is then
\begin{equation} 
\nu_{b\;QGP}(\beta,\lambda_u,\lambda_d)=
\frac{N_s N_c}{2\pi^2} \sum_i N_i \int_0^{\infty}
\frac{p^{2}}{e^{\beta \sqrt{p^2+m_i^2}}\lambda_i^{-1}+1}dp\;,
\end{equation}
where $i$ includes only $u$, $\bar{u}$, $d$ and $\bar{d}$ quarks and $N_i=1$
for $u$ and $d$ quarks and $N_i=-1$ for $\bar{u}$ and $\bar{d}$ quarks. The
pressure is
\begin{equation} 
P_{QGP}(\beta,\lambda_u,\lambda_d)=
\frac{1}{\beta}\frac{\ln Z_{QGP}(V,\beta,\lambda_u,\lambda_d)}{V}\;.
\end{equation}

In order to study the effect of the inclusion of strange quarks we can use
the partition function (11) and add the part of the quark term of (14) which
corresponds to the strange quarks.

\newpage
\vspace{2cm}
{\bf 4 Matching the two phases}

After completing a thermodynamic description for the hadronic and for the
quark-gluon phase we can trace whether a phase transition can occur between
the two phases. Similar situations have been studied in [10,12,13], but here,
apart from the use of the SBM incorporating all four improvements, we shall
focus our calculations to the location of the critical point. So no value of
$B$ or $a_s$ will be selected a-priori.

If $a_s$ and $\xi$ are fixed, then the only free parameter left
would be the MIT bag constant $B$. If a value of $B$ is chosen, also, the
pressure-volume isotherms of Hadron Gas and QGP can be calculated for a specific
temperature. Then at the point where the two isotherms meet would correspond
equal volumes and equal pressures for the two phases. But assuming that the
baryon number is a conserved quantity to both phases, the equality of volumes
would lead to the equality of baryon densities.

When performing calculations about the location of the point where the two
phases meet, with fixed MIT bag constant, what is found is that at a low
temperature the QGP and SBM pressure-volume isotherms meet at a point where
the Hadron Gas pressure is {\em decreasing} while volume {\em decreases}.
This is reminiscent of the need of a Maxwell construction. So at that point
the phase transition between Hadron Gas and QGP must be of {\em first order}. 
As the temperature rises, a certain temperature is found for which the QGP
isotherm meets the SBM isotherm at a point which corresponds to the maximum
Hadron Gas pressure for this temperature. So no Maxwell construction is
needed. It is important to notice that this point is located at {\em finite
volume} or {\em finite baryon density} and it can be associated with the
QCD critical point.
Then, as temperature continues to rise, the QGP isotherms meet the SBM
isotherms at points with even greater volume. Again no Maxwell construction is
needed and this region belongs to the {\em crossover} area.

These situations can be depicted in Figure 2, where all curves have been
calculated for $B^{1/4}=210$ MeV. The dotted curved lines correspond to SBM,
while the almost straight dotted lines correspond to QGP. For the
calculations three quark flavours have been used with their corresponding
current masses and $a_s=0$. The thick lines are the resulting pressure-volume 
curves for the Hadron Gas-QGP system. A Maxwell construction is needed for
the low temperature isotherm. This is depicted by the horizontal line
which is drawn so that the two shaded surfaces are equal and represents the 
final pressure-volume curve after the completion of the Maxwell construction. 
In the same figure the isotherm that leads the pressure curves of the two
phases to meet at the maximum hadron gas pressure, forming a critical point,
is drawn, also. Finally for higher temperatures the two curves meet at a
point so that the resulting pressure curve is always increasing as volume
decreases, without the need of a Maxwell construction (crossover area).

A more detailed figure of the previous one is Figure 3, where more curves
that need Maxwell construction can be displayed. The coexistence region of the
two phases are represented by the horizontal Maxwell constructed curves. The
slashed line represents the boundaries of the Maxwell construction and so the
boundaries of the coexistence region.

\vspace{2cm}
{\bf 5 Locating the Critical Point}

To locate the critical point with the choice (14) for the QGP partition
function, for a given $B$, one has to determine the parameters
$(\beta, \lambda_u, \lambda_d, \lambda_s, \lambda_u', \lambda_d')$, which
solve the following set of equations. 

\[
\hspace{4.2cm}
\nu_{b\;SBM}(\beta,\lambda_u,\lambda_d,\lambda_s)=
\nu_{b\;QGP}(\beta,\lambda_u',\lambda_d')
\hspace{2.7cm}(17{\rm a})
\]
\[
\hspace{4.3cm}
P_{SBM}(\beta,\lambda_u,\lambda_d,\lambda_s)=
P_{QGP}(\beta,\lambda_u',\lambda_d')
\hspace{2.9cm}(17{\rm b})
\]
\[
\hspace{5.5cm}
\frac{\partial P_{SBM}(\beta,\lambda_u,\lambda_d,\lambda_s)}
{\partial \lambda_u}=0
\hspace{4cm}(17{\rm c})
\]
\[
\hspace{5.4cm}
\left<\;S(\beta,\lambda_u,\lambda_d,\lambda_s)\;\right>_{SBM}=0
\hspace{3.9cm}(17{\rm d})
\]
\[
\hspace{2.8cm}
\left<\;b(\beta,\lambda_u,\lambda_d,\lambda_s)\;\right>_{SBM}-
2\;\left<\;Q(\beta,\lambda_u,\lambda_d,\lambda_s)\;\right>_{SBM}=0
\hspace{1.9cm}(17{\rm e})
\]
\[
\hspace{3.4cm}
\left<\;b(\beta,\lambda_u',\lambda_d')\;\right>_{QGP}-
2\;\left<\;Q(\beta,\lambda_u',\lambda_d')\;\right>_{QGP}=0
\hspace{2.5cm}(17{\rm f})
\]
Eq. (17c) is equivalent to $P_{SBM}=P_{SBM\;max}$, when all the rest of the
equations are valid. Eq. (17d) imposes zero strangeness to HG phase.
Eqs. (17e) and (17f) account for isospin symmetry in the HG and QGP phase,
respectively. Also we have set $\gamma_s=1$ assuming full strangeness
equilibrium.

With the choice (11) for the QGP partition function only the equations
(17a)-(17e) have to be solved, since only one fugacity
$\lambda_q=\lambda_u'=\lambda_d'$ is available in the QGP phase.

The calculations for the position of the critical point for different values
of $B$ are presented in Figures 4-6. The range of values of
$B^{1/4}=(145-235)$ MeV [14,15] has been used for these calculations. In
Figure 4 we depict the critical temperature as a function of the critical
baryon density.
The dotted curves correspond to the QGP partition function with massless $u$
and $d$ quarks, without strange quarks and for different values of $a_s$.
The thick solid curve corresponds to the QGP partition function with massive
$u$, $d$ and $s$ quarks and $a_s=0$.
The slashed curve corresponds to the QGP partition function with massless $u$,
$d$, massive $s$ quarks and $a_s=0.1$.

Figure 5 presents the connection of the MIT bag constant with the baryon
density of the critical point, divided by the normal nuclear density.

In Figure 6 the critical temperature is plotted versus the critical
baryon chemical potential. The code of lines are as in Figure 4.
In this graph the
lines representing the bootstrap singularity, that is the boundaries of the
maximum space allowed to the hadronic phase, for the maximum and minimum
values of $B$, are also depicted (slashed-dotted curves). The filled circles
represent positions of critical point for the different choices of the QGP
partition functions for these maximum and minimum values of $B$.
As it can be seen the critical point is placed within the hadronic phase,
close to the bootstrap singularity. Every modification made to external 
parameters drives the critical point in parallel to the bootstrap
singularity line.

Typical values for the position of the critical point are listed in Table 1.

\vspace{2cm}
{\bf 6 Concluding Remarks}

From our study we may conclude that, as $B$ increases, the critical point
moves to higher baryon density, smaller baryon chemical potential and higher
temperature until a certain value of $B$ is reached. If $B$ is increased
further, then the critical point moves quickly to zero baryon density and
zero baryon chemical potential, while temperature keeps increasing slowly.

The inclusion of strange quarks always moves the critical point to higher
baryon density and higher baryon chemical potential (for fixed values of $B$
and $a_s$).

As $a_s$ is increased (at the same QGP partition function), the critical point
moves to smaller baryon density, smaller baryon chemical potential and higher
temperature, while the move of the critical point towards zero chemical
potential takes place at smaller values of $B$.

From the last two remarks we can infer that the calculation with massive
quarks and $a_s=0$ represents the {\it higher} baryon density,
{\it higher} baryon chemical potential and {\it smaller} temperature (for a
given B) that the critical point can acquire. So this particular QGP partition
function can give us an {\it upper} limit for the position of the critical
point in baryon density or baryon chemical potential.

From Figure 6 it is evident that the critical point is positioned near the
bootstrap singularity curve. So this curve can represent, to a good
approximation, the first-order transition line between hadron and quark-gluon
phase.

From Table 1 we observe that in the minimal, two flavour version of the
quark-gluon description ($a_s=0$) and in the chiral limit ($m_u=m_d=0$),
where the critical point becomes tricritical, the location of the singularity
may come close to the freeze-out area of the SPS experiments (typically:
$T_c \approx 171$ MeV, $\mu_c \approx 300$ MeV). On the contrary, the Lattice
QCD solution [16] with unphysically large values of the quark masses $m_u$,
$m_d$ drives the critical baryon chemical potential to higher values
($T_c \approx 160$ MeV, $\mu_c \approx 725$ MeV). In order to bridge this
discrepancy one needs an improvement in both approaches. In the bootstrap
approach a realistic partition function of the quark-gluon matter is needed,
based not on perturbation theory but on the knowledge of the quark-gluon
pressure on the lattice for nonzero chemical potential. At present, there
exist lattice results for the pressure only for $\mu=0$ [17]. In the lattice
search for the critical point on the other hand the solution for small quark
masses (chiral limit) is needed before any quantitative comparison, both with
the bootstrap solution and the location of the freeze-out area in heavy-ion
collisions, could be made.

{\bf Figure Captions} 
\newtheorem{f}{Figure} 
\begin{f} 
\rm Isotherm pressure-volume curve for SBM and IHG (both with Van der Waals
    volume corrections using the pressure ensemble).  $B$ is constant.
\end{f} 
\begin{f} 
\rm Three isotherm pressure-volume curves for Hadron Gas (using SBM) and QGP
    phase (using partition function including $u$, $d$ and $s$ quarks at
    their current masses and $a_s=0$). The low temperature isotherm
    needs Maxwell construction, the middle temperature isotherm corresponds
    to critical point and the high temperature isotherm corresponds to
    crossover. $B$ is constant.
\end{f} 
\begin{f} 
\rm A similar case as in Figure 2. The boundaries of Maxwell construction
    are displayed with the slashed line.
\end{f} 
\begin{f} 
\rm The baryon density at the critical point versus the critical
    temperature for different values of $B$ and for different types of the QGP
    partition function.
\end{f} 
\begin{f} 
\rm The critical temperature as a function of the MIT bag constant
    for different types of the QGP partition function.
\end{f} 
\begin{f}
\rm Critical temperature versus critical baryon chemical
    potential for different values of $B$ and for different types of QGP
    partition functions. The bootstrap singularity lines for maximum and
    minimum values of $B$, as well as, the critical points corresponding to
    these values (filled circles) are also displayed.
\end{f}

\vspace{3cm}
\begin{center}
\begin{tabular}{|cccc|} \hline
$B^{1/4}$ (MeV) & $\nu_{b\;cr.p.}$ (fm$^{-3}$) & $T_c$ (MeV) &
$\mu_c$ (MeV) \\ \hline
\multicolumn{4}{|c|}{$a_s=0$, $m_u=m_d=0$, $s$-quarks not included} \\ \hline
235 & 0.2158 & 171.2 & 299.4 \\
180 & 0.1361 & 127.9 & 544.5 \\
145 & 0.0690 & 102.6 & 623.4 \\ \hline
\multicolumn{4}{|c|}{$a_s=0$, $m_u=5.6$ MeV, $m_d=9.9$ MeV, $m_s=199$ MeV} \\ \hline
235 & 0.3110 & 159.1 & 451.1 \\
180 & 0.1489 & 121.2 & 598.6 \\
145 & 0.0721 &  98.4 & 651.9 \\ \hline
\end{tabular}
\end{center}

\begin{center}
Table 1.
\end{center}

{\bf Table Caption} 
\newtheorem{g}{Table} 
\begin{g} 
\rm Some values for the position of the critical point for different values
    of $B$ and different QGP partition functions.
\end{g}

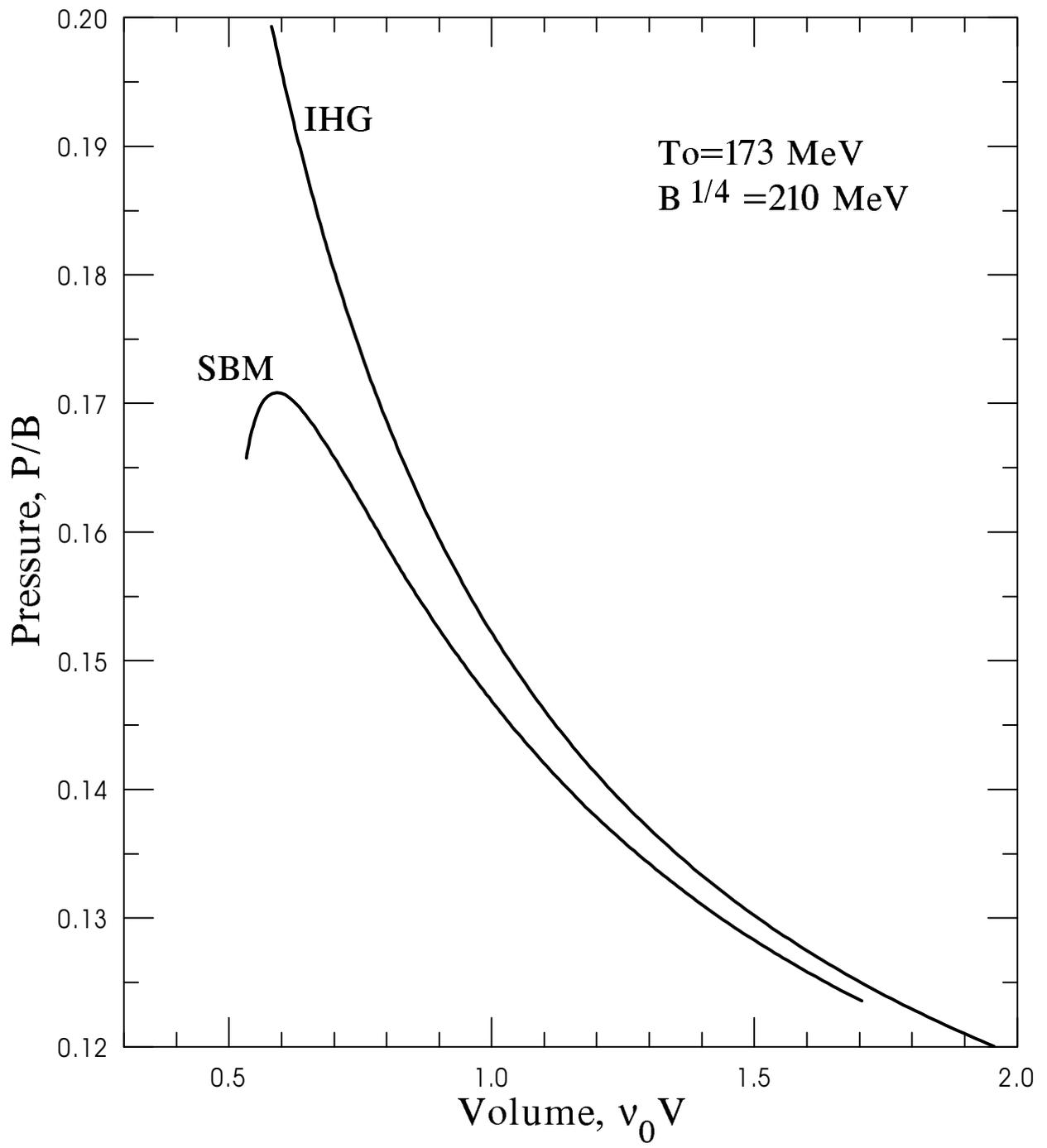
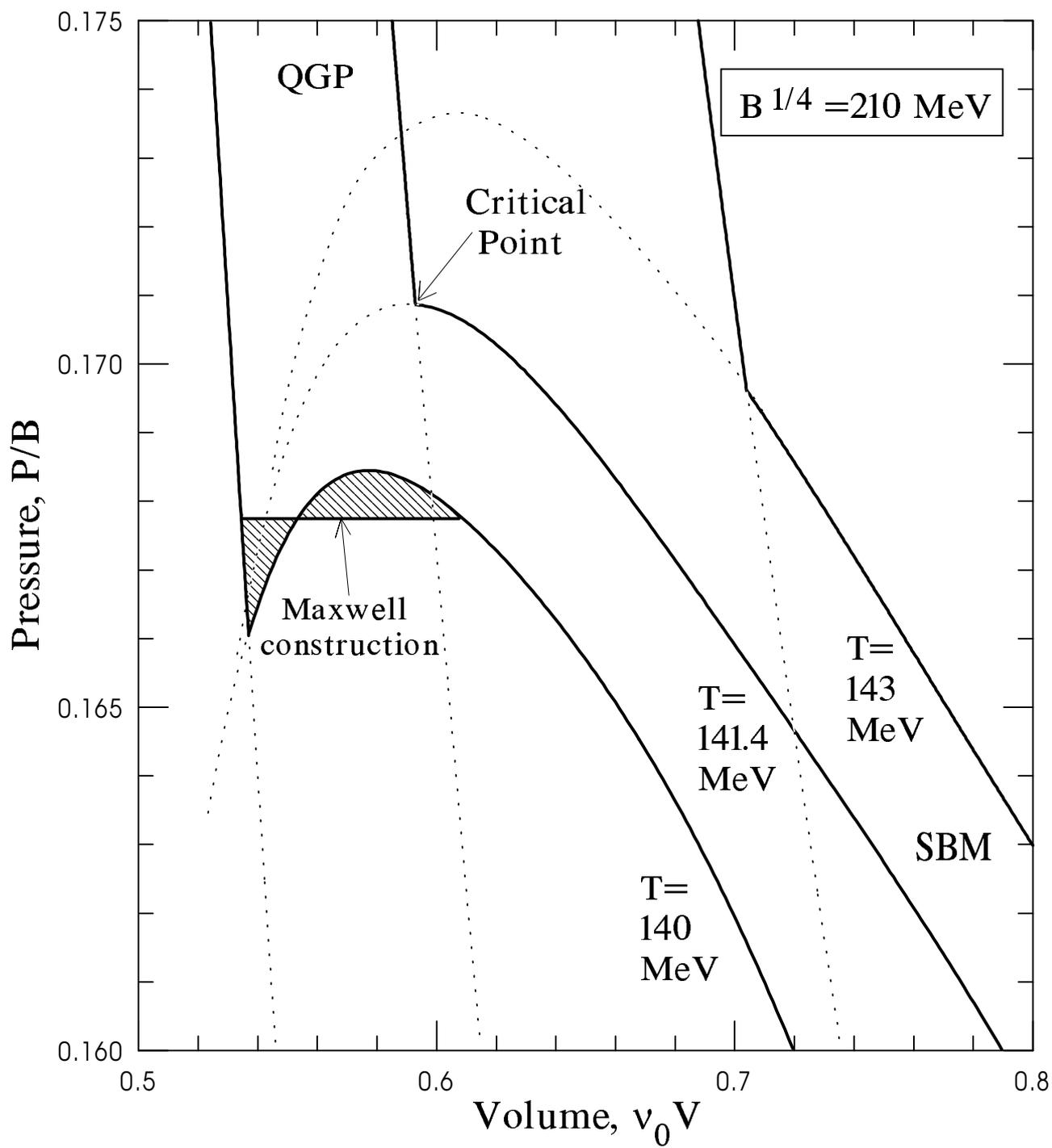
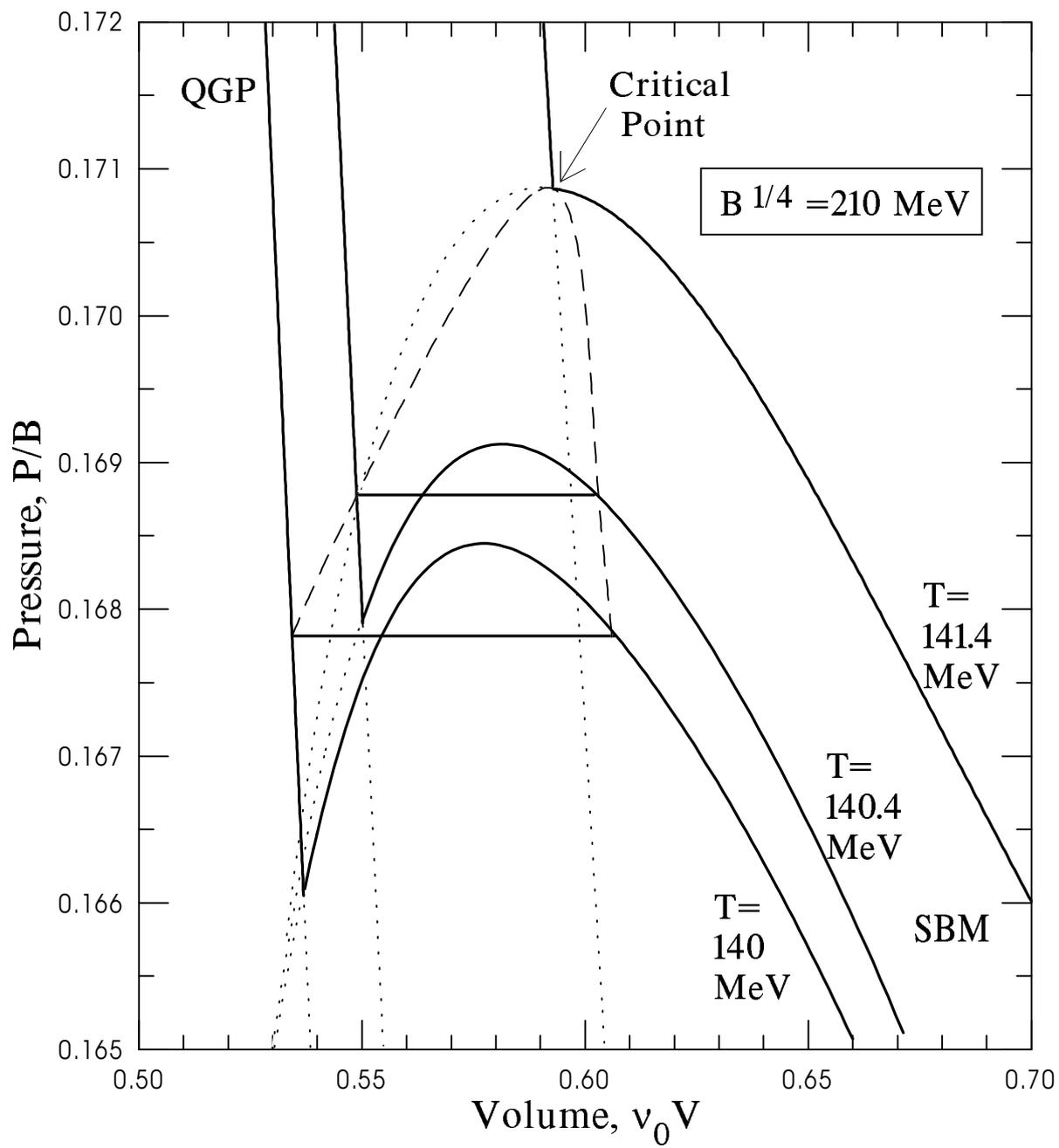
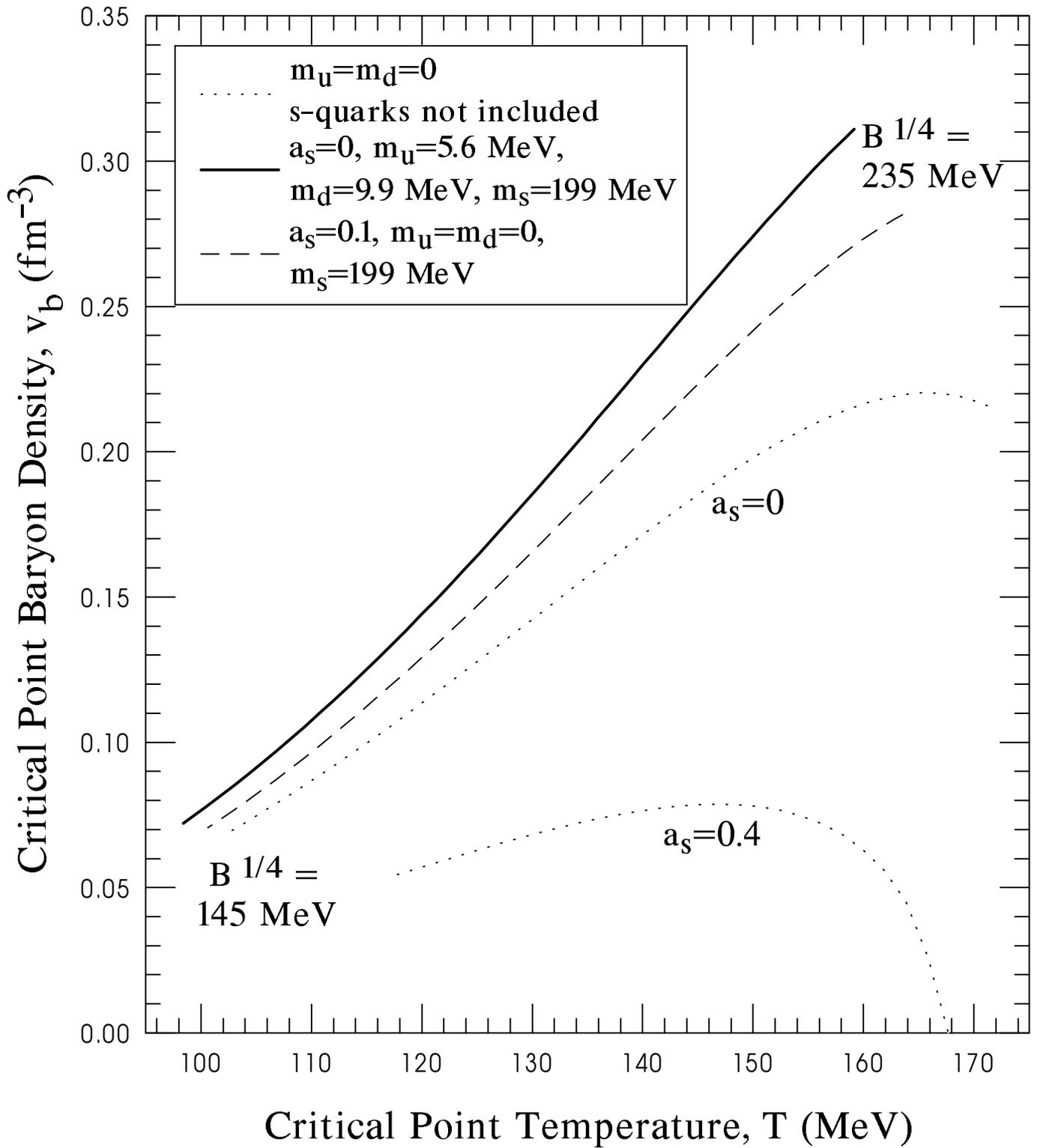
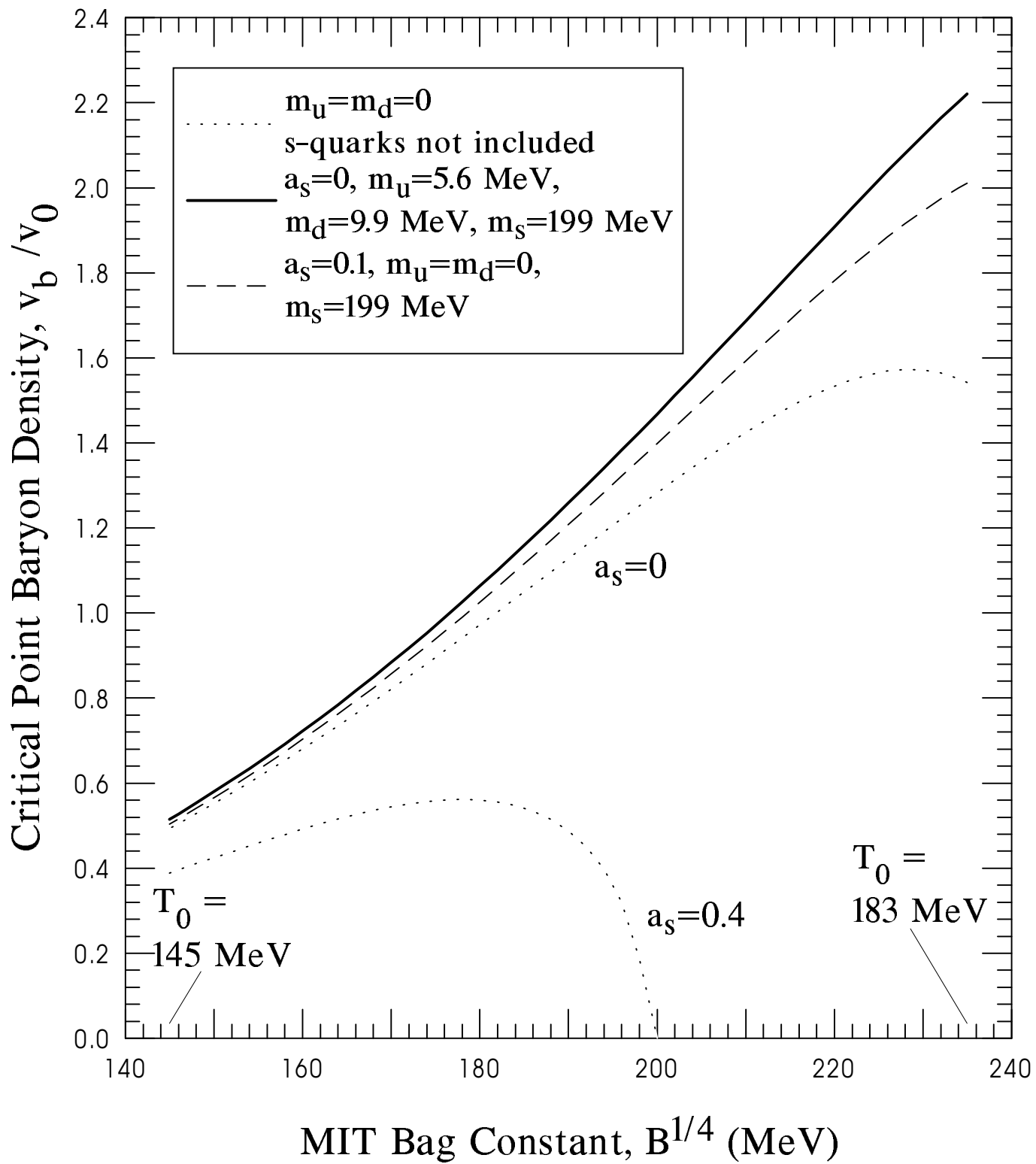
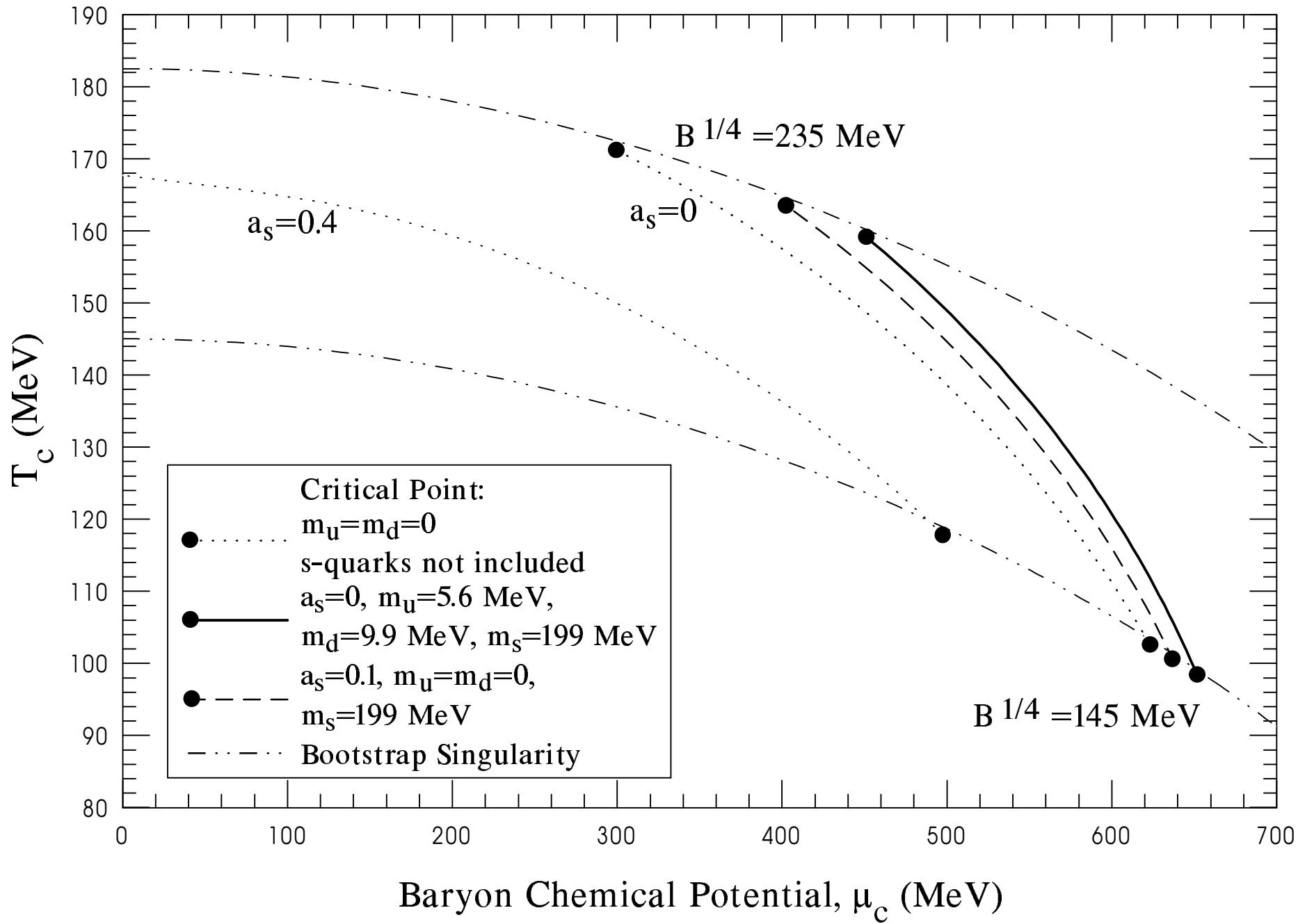
\end{document}